\documentclass[aps,prl,twocolumn,notitlepage,superscriptaddress,showpacs,10pt,longbibliography,floatfix]{revtex4-1} 

\usepackage{amsmath}    
\usepackage{amsfonts}
\usepackage{bm}
\usepackage{amssymb}
\usepackage{graphicx}   
\usepackage[usenames,dvipsnames]{xcolor}      
\usepackage{comment}
\usepackage{siunitx}
\usepackage{enumitem}
\usepackage{upgreek}

\usepackage{graphicx}

\usepackage{soul}
\usepackage[normalem]{ulem}

\mathchardef\mhyphen="2D

\newcommand{\fref}[1]{Fig.\,\ref{#1}}

\DeclareSIUnit[qualifier-mode = space]{\elec}{el}

\begin{document}

\title{Topological phenomena in artificial quantum materials revealed by local Chern markers}

\author{Catalin D. Spataru}
\email[]{cdspata@sandia.gov}
\affiliation{Sandia National Laboratories, Livermore, California 94551, USA}

\author{Wei Pan}
\email[]{wpan@sandia.gov}
\affiliation{Sandia National Laboratories, Livermore, California 94551, USA}

\author{Alexander Cerjan}
\email[]{awcerja@sandia.gov}
\affiliation{Center for Integrated Nanotechnologies, Sandia National Laboratories, Albuquerque, New Mexico 87185, USA}

\date{\today}

\begin{abstract}
A striking example of frustration in physics is Hofstadter's butterfly, a fractal structure that emerges from the competition between a crystal's lattice periodicity and the magnetic length of an applied field.
Current methods for predicting the topological invariants associated with Hofstadter's butterfly are challenging or impossible to apply to a range of materials, including those that are disordered or lack a bulk spectral gap. 
Here, we demonstrate a framework for predicting a material's local Chern markers using its position-space description and validate it against experimental observations of quantum transport in artificial graphene in a semiconductor heterostructure, inherently accounting for fabrication disorder strong enough to close the bulk spectral gap. 
By resolving local changes in the system's topology, we reveal the topological origins of antidot-localized states that appear in artificial graphene in the presence of a magnetic field.
Moreover, we show the breadth of this framework by simulating how Hofstadter's butterfly emerges from an initially unpatterned 2D electron gas as the system's potential strength is increased, and predict that artificial graphene becomes a topological insulator at the critical magnetic field.
Overall, we anticipate that a position-space approach to determine a material's Chern invariant without requiring prior knowledge of its occupied states or bulk spectral gaps will enable a broad array of fundamental inquiries and provide a novel route to material discovery, especially in metallic, aperiodic, and disordered systems.
\end{abstract}

\maketitle

Over the last half-century, few physical systems have been studied as intently as two-dimensional electron gasses subjected to a perpendicularly oriented magnetic field. By itself, this configuration yields the integer quantum Hall effect, whose defining feature is a quantized conductivity stemming from protected edge-localized transport channels \cite{klitzing_new_1980,laughlin_quantized_1981,halperin_quantized_1982}. However, when a periodic electrostatic potential is applied, the system becomes frustrated---the applied magnetic field $B$ is attempting to drive the system to exhibit degenerate Landau levels, characterized by the magnetic length $l_B = \sqrt{\hbar/eB}$, while the periodic potential tries to force the system to exhibit extended Bloch modes, characterized by the lattice constant $a$. This competition results in each band being split into subbands separated by minigaps that form a fractal as a function of the applied magnetic field and Fermi energy, Hofstadter's butterfly \cite{hofstadter_energy_1976}. The fractal structure for a spectrally isolated underlying band is periodic in $\Phi / \Phi_0$, where $\Phi = BA$ is the magnetic flux through a unit cell with area $A$ and $\Phi_0 = h/e$ is the magnetic flux quantum. 
Hofstadter's butterfly has been observed in artificial quantum materials \cite{schlosser_landau_1996,albrecht_evidence_2001,geisler_detection_2004} and more recently in graphene superlattices \cite{dean_hofstadters_2013,hunt_massive_2013,ponomarenko_cloning_2013,forsythe_band_2018,lu_multiple_2021}.

Despite substantial progress in achieving system periodicities large enough to enter the parameter regime where Hofstadter's butterfly manifests, $\Phi \sim \Phi_0$, it remains a formidable challenge to predict the Chern numbers associated with each minigap in experimentally realizable systems. For low-energy models with limited degrees of freedom, there are a few different approaches to predicting a minigap's Chern number: via a Diophantine equation \cite{wannier_result_1978,macdonald_quantized_1984,dana_quantised_1985,sato_hall_2008,avron_study_2014}, St{\v r}eda's formula \cite{streda_quantised_1982}, semi-classical analysis \cite{chang_berry_1996}, bulk-boundary correspondence \cite{hatsugai_topological_2006,avila_topological_2013,agazzi_colored_2014}, or direct calculation using the occupied states \cite{thouless_quantized_1982,kitaev2006anyons,bianco_local_chern_2011}. However, these methods are impossible or impractical to apply to many experimental platforms, stymied either by prohibitively large computational costs in the absence of a low-energy description, the lack of a bulk spectral gap due to disorder,  or the need for system-specific knowledge \cite{avron_study_2014,lian_open_2021,janecek_two-dimensional_2013}. In such cases, the last resort is direct simulation of a system's quantum transport \cite{groth_kwant_2014,de_castro_fast_2024}, requiring the specification of a device geometry and still yielding a costly computational endeavor for realistic systems. Moreover, such an approach will miss bulk-embedded phenomena that do not contribute to the chosen transport channels.

Here, we theoretically demonstrate and experimentally validate the spectral localizer framework \cite{loringPseuspectra,LoringSchuBa_odd,LoringSchuBa_even} for predicting a quantum material's local Chern topology and associated boundary-localized states. Moreover, we show how the spectral localizer framework can be used to reveal distinct material topology at different length scales of multi-scale systems.
Using this framework in artificial graphene \cite{park_electron_2008,wunsch_dirac-point_2008} subjected to an out-of-plane magnetic field and described by a continuum model without a low-energy approximation, we demonstrate quantitative agreement between the Chern marker and the experimentally observed Hall conductivity, while inherently accounting for fabrication disorder that is strong enough to remove the system's spectral gaps. Moreover, by spatially resolving local changes in the system's topology, we show that many of the pinned states that populate the gaps between the bulk Landau levels are topological, stemming from the distinct topology of the antidots relative to the unpatterned bulk for many magnetic field strengths.
Taking advantage of the spectral localizer's ability to efficiently operate without a low-energy approximation, we numerically observe the formation of Hofstadter's butterfly from an unpatterned system. Finally, we 
predict the opening of a topological band gap in artificial graphene at the critical magnetic field due to long-range couplings, yielding
a topological insulator at modest magnetic fields. Looking forward, we anticipate that the spectral localizer framework's application to realistic multi-scale systems will both yield a novel approach to material classification and enable a broad array of fundamental inquiries, such as those into the formation of Hofstadter's butterfly in twisted materials \cite{bistritzer_moire_2011,hejazi_landau_2019,zhang_landau_2019}, as well as aid in the search for materials that exhibit the quantum anomalous Hall effect \cite{nagaosa_anomalous_2010,chang_experimental_2013,sharpe_emergent_2019,deng_quantum_2020,serlin_intrinsic_2020,chen_tunable_2020,li_quantum_2021,zhao_realization_2024}.

\begin{figure}[t]
    \centering
    \includegraphics{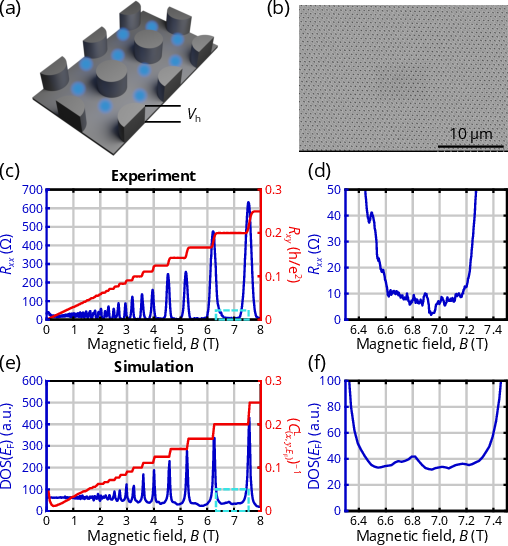}
    \caption{(a) Schematic of the antidot potential profile (gray) with strength $V_{\textrm{h}}$ and locations where the electrons are approximately localized (blue) to form the artificial graphene pseudo-atoms. (b) SEM image of a representative experimental specimen. (c) Measured longitudinal $R_{xx}$ (blue) and transverse $R_{xy}$ (red) resistances versus the magnetic field in the artificial graphene AlSb/InAs/AlSb semiconductor heterostructure at $\SI{0.3}{\kelvin}$ with lattice constant $a=\SI{250}{\nm}$, antidot diameter $a/2$, and electron density $n \sim \SI{8e11}{\per \cm\squared}$. (d) Zoom-in of the measured $R_{xx}$ in the dashed cyan box in (c).
    (e) Calculated density of states (DOS, blue) and $(C_{(x,y,E_\textrm{F})}^{\textrm{L}})^{-1}$ calculated in the unpatterned bulk versus the applied magnetic field at the Fermi energy $E_\textrm{F} = \SI{85}{\meV}$ for artificial graphene with the same geometry as the experimental system with simulation flake size $\sim2\times3.4~\SI{}{\micro\meter^2}$, and model parameters $V_h=\SI{25}{\meV}$, normally distributed disorder with standard deviation $\delta V_\textrm{h}=\SI{5}{\meV}$ per antidot, $m^*=0.023 m_0$, $g=40$, and discretization $\Delta x =\SI{2}{\nm}$. $(C_{(x,y,E_\textrm{F})}^{\textrm{L}})^{-1}$ simulations use $\kappa = \SI{1E-3}{\meV/\nm}$. (f)  Zoom-in of the simulated $\textrm{DOS}(E_\textrm{F})$ in the dashed cyan box in (e).}
    \label{fig:exp-L250}
\end{figure}

To motivate the development of our theoretical approach, we consider an electron gas confined to an effectively 2D quantum well layer in InAs surrounded by barrier layers of AlSb in a semiconductor heterostructure, into which an antidot triangular lattice is added via interferometric lithography. The triangular antidot lattice confines the electron gas in-plane to areas furthest from the antidots yielding an effective honeycomb lattice for the electrons in the low-potential regions between three antidots coupled together via the potential troughs between two antidots [Fig.\ \ref{fig:exp-L250}a,b], altogether creating artificial graphene.
It has been shown \cite{park_electron_2008,wunsch_dirac-point_2008,wang_observation_2018,du_emerging_2018,cheng_majorana_2020,tkachenko_wannier_2022} that the low-energy electronic band structure of artificial graphene mimics the behavior of natural graphene's Dirac cones. 
In the presence of a perpendicular, static magnetic field $\mathbf{B} = B \hat{\mathbf{z}}$, the 2D electron gas in artificial graphene is assumed to be non-interacting and characterized by the single-particle Hamiltonian
\begin{equation}
    H = \frac{1}{2m^*} \left(-i \hbar \nabla + e\mathbf{A}(\mathbf{x}) \right)^2 + V(\mathbf{x}) - \frac{\mu_B g}{\hbar} s_z B, \label{eq:Hcont}
\end{equation}
where the electrons have effective mass $m^*=m_{\textrm{eff}} m_0$, $V(\mathbf{x})$ is a scalar potential that accounts for the system's nanoscale structure, and the strength of the Zeeman splitting is proportional to the effective Land{\' e} $g$-factor of the heterostructure and the electron's spin $s_z$.

The measured longitudinal ($R_{xx}$) and Hall ($R_{xy}$) resistivities of our AlSb/InAs/AlSb artificial graphene heterostructure are shown in Figs.~\ref{fig:exp-L250}c,d. Overall, the $R_{xx}$ and $R_{xy}$ traces resemble those of a typical 2D electron gas in an unpatterned heterostructure. Shubnikov de-Haas oscillations (SdH) are observed at low magnetic field strengths, and a Fast-Fourier Transform analysis of these SdH oscillations yields an electron density of $n \sim \SI{8e11}{\per \cm\squared}$. In the high $B$-field regime, fully developed quantized Hall states are formed, with $R_{xx}$ assuming a low resistance value and $R_{xy}$ quantized to the value of $h/e^2\nu$ in between Landau levels with filling factor $\nu=nh/eB$. 

To numerically model artificial graphene using a position-space description, we first consider the zero-field limit of Eq.~(\ref{eq:Hcont}) and approximate the Laplacian in the kinetic term $K = -(\hbar^2/2m^*) \nabla^2$ using finite-difference methods \cite{thomas_numerical_1995} with vertex spacings $\Delta x = \Delta y$.
This procedure transforms $K$ into a sparse bounded matrix $K^{(\textrm{FD})}$ in which adjacent vertices in the square lattice are coupled together with strength $t^{(\textrm{FD})}= \hbar^2 /(2m^* \Delta x^2)$. For electron energies $E$ sufficiently smaller than $t^{(\textrm{FD})}$, 
$K^{(\textrm{FD})}$ accurately describes a free electron.
When discretized, the potential energy $V(\mathbf{x})$  becomes a diagonal matrix $V^{(\textrm{FD})}$ representing the potential strength at each vertex.
Finally, the magnetic field is re-introduced  both by using the Peierls substitution \cite{graf_electromagnetic_1995} in $K^{(\textrm{FD})}$ (see Supplemental Sec.~SI \cite{SI}), yielding a $B$-dependent phase to some of the couplings $t^{(\textrm{FD})}$, and by including the Zeeman splitting in $V^{(\textrm{FD})}$.
Our model is parameterized by comparing $R_{xx}$ against the density of states (DOS) of the discretized system [Figs.~\ref{fig:exp-L250}e,f], finding a Fermi level $E_F\equiv \pi n\hbar^2/m^{\textrm{*}}\sim \SI{85}{\meV}$, and $g=40$ using a uniform Zeeman splitting approximation \cite{krishtopenko_theory_2011,pan_quantitative_2011}.

However, in contrast to a free 2D electron gas, careful examination of $R_{xx}$ in Figs.~\ref{fig:exp-L250}c,d shows additional features in between the system's main Landau levels, e.g., near $B = \SI{7}{\tesla}$. These features are numerically reproduced by choosing a disordered antidot potential strength $V_\textrm{h} + \delta V_\textrm{h} \xi$, with $V_\textrm{h}=\SI{25}{\meV}$, $\delta V_\textrm{h}= \SI{5}{\meV}$, and normally distributed $\xi$. Simulations show that the between-Landau level features correspond to states at the Fermi energy pinned to the antidots [Fig.\ \ref{fig:antidot}] whose fine features in the ordered DOS are blurred out by disorder (see Supplemental Sec.~SII \cite{SI}). As such, the difference in the relative prominence of these features in the observed $R_{xx}$ versus the simulated DOS can be understood as these states' spatial pinning limiting their charge mobility and thus limiting their contribution to 
$R_{xx}$ \cite{altshuler_magnetoresistance_1980,weiss_electron_1991}. Experimentally, we also see that these pinned states have a vanishing contribution to the Hall conductance. Yet, as these antidot-localized states completely fill the bulk spectral gap, they prohibit the use of many approaches for predicting the Hall conductivity. 

\begin{figure}[t]
    \centering
    \includegraphics{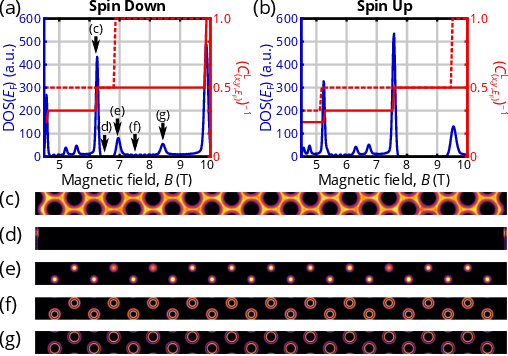}
    \caption{(a,b) Calculated DOS (blue) and local Chern marker inverse (red) for an ordered version of the system considered in Fig.~\ref{fig:exp-L250}, separated by spin. The local Chern marker is calculated both in the unpatterned bulk (solid) and center of an antidot (dashed) using $\kappa = \SI{0.5}{\meV/\nm}$. (c-g) LDOS of the spin down sector at $E_{\textrm{F}}$ showing bulk Landau levels at $B = \SI{6.21}{\tesla}$ (c), chiral edge states at $B = \SI{6.49}{\tesla}$ (d), antidot-localized Landau levels at $B = \SI{6.94}{\tesla}$ (e), and chiral antidot-bulk interface-localized states at $B = \SI{7.52}{\tesla}$ (f) and $B = \SI{8.42}{\tesla}$ (g).}
\label{fig:antidot}
\end{figure}

Instead, to predict the Chern invariants between the Landau levels of disordered artificial graphene despite the lack of a spectral gap at the Fermi energy, we employ the spectral localizer framework \cite{loringPseuspectra,LoringSchuBa_odd,LoringSchuBa_even} that has previously been successful at classifying the topology of gapless acoustic metamaterials \cite{Cheng2023}, photonic crystals \cite{cerjan_operator_Maxwell_2022,dixon_classifying_2023}, and toy models \cite{cerjan_local_2022}. The spectral localizer is a composite operator formed by combining the eigenvalue equations of a finite system's Hamiltonian and position operators using a Clifford representation; in 2D, the Pauli matrices can be used, yielding
\begin{multline}
    L_{(x,y,E)}(X,Y,H) = \kappa (X-x{\bf 1})\otimes \sigma_x \\ +  \kappa (Y-y{\bf 1})\otimes \sigma_y + (H-E{\bf 1})\otimes \sigma_z. \label{eq:L}
\end{multline}
Here, $X$ and $Y$ are position operators, ${\bf 1}$ is the identity, and $\kappa > 0$ is a scaling coefficient that ensures consistent units and similar spectral weighting between the summands. Heuristically, the choice of $\kappa$ is analogous to the choice of integration region necessary for other local Chern markers \cite{kitaev2006anyons,bianco_local_chern_2011}. 
The approximate scale of $\kappa$ is set by the smallest dimension of the finite system $L_{\textrm{min}}$ and the width of the relevant bulk spectral gap $E_{\textrm{gap}}$, as $\kappa \sim E_{\textrm{gap}}/L_{\textrm{min}}$ \cite{cerjan_local_crystal_marker_2024}. For artificial graphene in the absence of electron interactions, the Landau level spacing sets the size of the spectral gap, $E_{\textrm{gap}}(B) = \hbar \omega_\textrm{c}(B)$ where $\omega_\textrm{c}(B)$ is the cyclotron frequency. In practice, choices of $\kappa$ spanning many orders of magnitude provide quantitatively similar results (see Supplemental Sec.~SIII \cite{SI}), though increasingly large simulation domains are needed for weak magnetic fields as $E_{\textrm{gap}}(B)$ decreases.

The 2D spectral localizer defines a local Chern marker
\begin{equation}
    C_{(x,y,E)}^{\textrm{L}}(X,Y,H) = \frac{1}{2} \textrm{sig} \left[L_{(x,y,E)}(X,Y,H) \right] \in \mathbb{Z}, \label{eq:CL}
\end{equation}
where $\textrm{sig}$ denotes the matrix's signature, its number of positive eigenvalues minus its number of negative ones \cite{loringPseuspectra,LoringSchuBa_odd,LoringSchuBa_even}.
Heuristically, the ability of Eqs.~\eqref{eq:L} and \eqref{eq:CL} to predict a system's Hall conductivity can be understood as follows: First, for a given choice of $(x,y,E)$, the spectral localizer is performing dimensional reduction from 2D to 0D, i.e., $L_{(x,y,E)}$ can be viewed as the Hamiltonian of a fictitious 0D system. Then, $C_{(x,y,E)}^{\textrm{L}}$ is calculating the 0th Chern number of this fictitious 0D system. Finally, because the dimensional reduction is consistent with Bott periodicity \cite{stone_symmetries_2010}, $C_{(x,y,E)}^{\textrm{L}}$ is equivalent to a local 1st Chern marker of the original 2D system at $(x,y,E)$. For an infinite, gapped system with $E$ in the relevant band gap, $C_{(x,y,E)}^{\textrm{L}}$ is provably equal to the 1st Chern number; in the thermodynamic limit the local Chern marker is defined through the spectral flow of $L_{(x,y,E)}$~\cite{LoringSchuBa_even,cerjan_classifying_2024}.

\begin{figure*}[tb]
    \centering
    \includegraphics{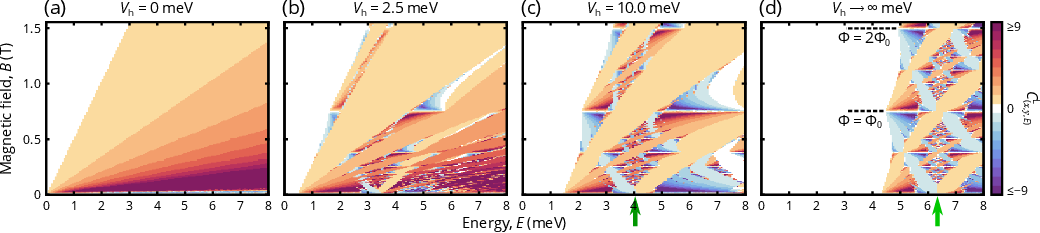}
    \caption{(a-d) Predicted $C_{(x,y,E)}^{\textrm{L}}$ in the unpatterned bulk for a single spin sector as a function of magnetic field strength and energy for ordered artificial graphene with lattice constant $a=\SI{80}{\nm}$, antidot diameter $a/2$, $m^*=0.030 m_0$, $g=0$. The flake size is $\sim2\times2.2~\SI{}{\micro\meter^2}$, with $\Delta x =\SI{2}{\nm}$ and $\kappa =\SI{4E-5}{\meV/\nm}$. The potential strength is increased from $V_{\textrm{h}} = \SI{0}{\meV}$ (a) to $\SI{2.5}{\meV}$ (b), $\SI{10}{\meV}$ (c), and $\infty \si{\meV}$ (d). Green arrows denote the Dirac point energy at $B=\SI{0}{\tesla}$ (see Supplemental Sec.~SIV \cite{SI}).
    }
\label{fig:L80Butterflies}
\end{figure*}

Choosing $(x,y,E_{\textrm{F}})$ in the unpatterned bulk and at the Fermi energy, we find quantitative agreement between the measured $R_{xy}$ and simulated $(C_{(x,y,E_{\textrm{F}})}^{\textrm{L}})^{-1}$ of a system with disordered antidots, despite the lack of a spectral gap at the Fermi energy at every magnetic field strength [Figs.~\ref{fig:exp-L250}c,e].
However, by choosing $(x,y,E_{\textrm{F}})$ within an antidot and increasing $\kappa \sim 2 E_{\textrm{gap}} /a$ to resolve phenomena at a smaller spatial scale corresponding to the antidot diameter $a/2$ (see Supplemental Sec.~SIII \cite{SI}), the local Chern marker reveals the topological origin of many of the antidot-pinned states. In particular, noting that $E_{\textrm{F}} > V_h$, the spectral localizer framework identifies that in an ordered system at large magnetic fields, pinned states can both form highly degenerate antidot-confined Landau levels across which the local Chern marker changes its value independent from the marker evaluated in the unpatterned bulk, as well as chiral states localized to the interface between the antidot and the unpatterned bulk when the two regions have different local Chern numbers [Fig.~\ref{fig:antidot}]. The distinction between these two types of pinned states can be seen in their local density of states (LDOS), see Fig.~\ref{fig:antidot}e-g, and their localization confirms why neither type of pinned state strongly contributes to the experimentally accessible $R_{xx}$ and $R_{xy}$. When the local topology of the antidot regions matches that of the unpatterned bulk, the LDOS reveals those edge conduction channels responsible for $R_{xy}$, see Fig.~\ref{fig:antidot}d; note, combining both spin sectors and adding disorder removes these $B$-field ranges where only chiral edge states exist to recover the DOS of Fig.~\ref{fig:exp-L250}c-f.

The ability to predict an experimental system's minigap Chern markers without needing to find its spectral gaps and occupied states, develop a $V_{\textrm{h}}$-customized effective model, or specify a transport geometry, offers a variety of possibilities. In Figs.~\ref{fig:L80Butterflies}a-d, we explore the emergence of Hofstadter's butterfly for a honeycomb lattice from an unpatterned 2D electron gas as the artificial graphene antidot potential strength is increased. In particular, by using a discretized version of the continuum Hamiltonian that automatically incorporates higher-energy phenomena, the spectral localizer framework can inherently consider the zero potential limit. As can be seen, for any positive potential strength, horizontal line segments with vanishing Chern markers, and about which the Chern marker changes sign, immediately appear for $\Phi/\Phi_0 \in \mathbb{Z}$. In the limit of $V_\textrm{h} \rightarrow \infty \si{\meV}$, Hofstadter's butterfly becomes nearly periodic about these lines as expected. These horizontal lines also persist at much larger $E_\textrm{F}$ than the energy of the Dirac point, potentially aiding in experimental design (see Supplemental Sec.~SV \cite{SI}). Additionally, our simulations also reveal how the minigaps in artificial graphene close and reopen as the antidot potential strength is increased so that the Chern invariant can change; for example, showing how the $C^{\textrm{L}} = -1$ minigap forms at the Dirac point in artificial graphene's low-energy bands at $B=\SI{0}{\tesla}$ and slowly supersedes the $C^{\textrm{L}} \ge 2$ minigaps as $V_\textrm{h}$ increases.



\begin{figure}[tb]
    \centering
    \includegraphics{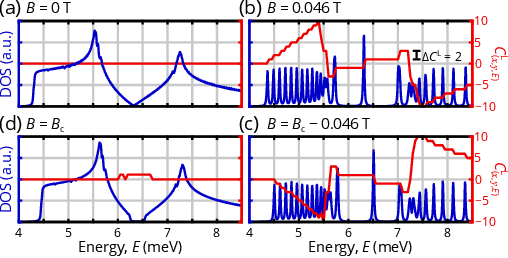}
    \caption{(a-d) Density of states (blue) and $C_{(x,y,E)}^{\textrm{L}}$ in the unpatterned bulk (red) for a single spin sector of artificial graphene with $V_{\textrm{h}} \rightarrow \infty \si{\meV}$ using the same simulation parameters as \fref{fig:L80Butterflies} for magnetic field strengths near $B \sim \SI{0}{\tesla}$ and the critical field $B \sim B_\textrm{c}$ where $\Phi = \Phi_0$.}
\label{fig:L80DOSind}
\end{figure}

One of the characteristic features of graphene subjected to a magnetic field is the manifestation of the unconventional quantum Hall effect for Fermi energies near the Dirac point \cite{zheng_hall_2002,gusynin_unconventional_2005,zhang_experimental_2005,novoselov_two-dimensional_2005-1,sheng_quantum_2006}. For sufficiently small $B$, this unconventional behavior is distinguished by a Landau level energy spacing following $E_j=v_\textrm{g}\sqrt{2e\hbar B |j|}$ with level index $j \in \mathbb{Z}$ and group velocity $v_\textrm{g}\approx h/(3 m^* a)$ in the vicinity of the Dirac point \cite{park_electron_2008}. Moreover, the Chern marker changes by $2$ per spin across each such Landau level. For weak magnetic field strengths $B \gtrsim \SI{0}{\tesla}$, the spectral localizer framework reproduces this behavior for artificial graphene, identifying that the Chern marker changes by $2$ per spin in the unconventional regime and the Landau level spacing is proportional to $\sqrt{|j|}$ (Figs.~\ref{fig:L80DOSind}a,b and Supplemental Sec.~SVI \cite{SI}). Given the periodicity of Hofstadter's butterfly, the spectral localizer framework also predicts the re-appearance of the unconventional quantum Hall effect near $B_\textrm{c}$ where $\Phi = \Phi_0$, with similar Landau level spacing but the opposite Hall conductivities for $B \lesssim B_\textrm{c}$ (Figs.~\ref{fig:L80DOSind}c,d and Supplemental Sec.~SVI \cite{SI}).

For a honeycomb lattice with only nearest neighbor (NN) couplings, Hofstadter's butterfly is perfectly periodic at $B = B_\textrm{c}$ where its DOS returns to that at $B=\SI{0}{\tesla}$. However, artificial graphene exhibits longer-range couplings between its pseudo-atoms as well whose effects are automatically incorporated through the use of a discretized continuum model; the approximate strength $t_{\textrm{NNN}}/t_{\textrm{NN}} \approx 0.13$  of the next-nearest neighbor (NNN) couplings can be estimated from the degree of chiral symmetry breaking in the $B=\SI{0}{\tesla}$ band structure (see Supplemental Sec.~SVI \cite{SI}). In the presence of a magnetic field, longer-range couplings alter the structure of Hofstadter's butterfly \cite{hou_NNN-tunneling-induced_2009}, and in artificial graphene they open a spectral gap around the $B=\SI{0}{\tesla}$ Dirac point that the spectral localizer predicts is topological $C_{(x,y,E)}^{\textrm{L}}=1$ [Fig.~\ref{fig:L80DOSind}d]. Thus, through careful control over the magnetic field in artificial graphene whose unit cell is large enough to yield experimentally accessible $B_\textrm{c}$, the spectral localizer framework shows that artificial graphene can become a topological insulator, offering opportunities for device applications \cite{novoselov_mind_2007}.


In conclusion, we have demonstrated how the spectral localizer framework can identify the topological origins of artificial graphene's antidot-localized states, shown the emergence of Hofstadter's butterfly across both the zero-potential and strong-potential limits, and predicted that artificial graphene becomes a topological insulator at $B_\textrm{c}$. A sample implementation of the spectral localizer is provided as part of the Supplemental Materials \cite{SI,spataru_github}. Overall, the spectral localizer framework possesses three key advantages: it can be applied directly to a material's single-particle position-space description without requiring a low-energy approximation, it can be applied without needing to find a system's occupied states or ensure the system has a bulk spectral gap at the Fermi energy, and it can reveal phenomena at different length scales of multi-scale systems. Looking forward, we expect that the spectral localizer framework can be applied to any weakly correlated material, including metallic and aperiodic materials, and thus offers an entirely distinct approach to topological material classification than existing methods that are based on a material's band structure \cite{bradlyn_topological_2017}.

\begin{acknowledgments}
We thank Terry A.\ Loring, Sachin Vaidya, and Joseph J.\ Cuozzo for helpful discussions.
The authors acknowledge support from the Laboratory Directed Research and Development program at Sandia National Laboratories. This work was performed, in part, at the Center for Integrated Nanotechnologies, an Office of Science User Facility operated for the U.S. Department of Energy (DOE) Office of Science. 
Sandia National Laboratories is a multimission laboratory managed and operated by National Technology \& Engineering Solutions of Sandia, LLC, a wholly owned subsidiary of Honeywell International, Inc., for the U.S.\ DOE's National Nuclear Security Administration under contract DE-NA-0003525. The views expressed in the article do not necessarily represent the views of the U.S.\ DOE or the United States Government.
\end{acknowledgments}

\bibliography{predicting_butterfly_refs}

\end{document}


\title{Supplemental Material for Topological phenomena in artificial quantum materials revealed by local Chern markers}

\author{Catalin D. Spataru}
\email[]{cdspata@sandia.gov}
\affiliation{Sandia National Laboratories, Livermore, California 94551, USA}

\author{Wei Pan}
\email[]{wpan@sandia.gov}
\affiliation{Sandia National Laboratories, Livermore, California 94551, USA}

\author{Alexander Cerjan}
\email[]{awcerja@sandia.gov}
\affiliation{Center for Integrated Nanotechnologies, Sandia National Laboratories, Albuquerque, New Mexico 87185, USA}

\maketitle

\section{Methods}
\subsection{Experiment}
Interferometric lithography (IL) is used to fabricate artificial graphene (AG). First, a square specimen of size \SI{5}{\milli\meter} $\times$ \SI{5}{\milli\meter} is cleaved from an as-grown AlSb/InAs/AlSb quantum well (QW) wafer. Then, a layer of photoresist of $\sim \SI{1.5}{\micro\meter}$ thick is spun on the specimen. To create the honeycomb structure needed for AG, interference of two coherent laser beams defines the first set of parallel lines. Then the specimen is rotated by $60^\circ$ and the second set of parallel lines is defined. The crossing points of the two sets of lines then create a triangular holes array after photo-resist develop. Subsequent reactive ion etching produces holes array in the 2D electron gas in the InAs QW. The unetched area forms a honeycomb lattice, leading to the formation of AG. An SEM (scanning-electron-microscope) image of a fabricated device, made using this procedure, is shown in Fig.~1b of the main text.

\subsection{Modeling}
We used the lattice parameters $\Delta x = \Delta y = \SI{2}{\nm}$ which together with $m_{\textrm{eff}}=0.023$ (typical of the experimental sample) yields $t^{(\textrm{FD})} \approx \SI{414}{\meV}$. This implies that the physics of the 2D electron gas is well described 
for $E_{\textrm{F}}$ within \SI{100}{\meV} from the bottom of the electron band. When a magnetic field is present, the finite different description is also only valid so long as the magnetic length is larger than the grid spacing, $l_B \gg \Delta x$. The potential $V(\mathbf{x})$ accounts for circular patterns via a muffin-tin on-site potential with strength $V_\textrm{h}$ and triangular symmetry. The AG pseudo-atoms develop with a honeycomb symmetry in between the circles (see Fig.~1a of the main text). 

A perpendicular magnetic field $\vec{B}=B\hat{\vec{z}}$ is introduced via the Peierls substitution within the Landau gauge $\vec{A}=B x \hat{\vec{y}}$ where $\vec{A}$ is the vector potential. In this gauge the hopping parameters $t^{(\textrm{FD})}$ between neighboring lattice sites $j,l$ take the form (for sites aligned along $\hat{\vec{x}}$ and $\hat{\vec{y}}$ respectively): $t_{j,l}^{(x)}=t$ and $t_{j,l}^{(y)}=t e^{\pm 2 \pi i n p /N_x}$ where $N_x$ is the number of vertices along $\hat{\vec{x}}$, $n\Delta_x$ is the $j$-vertex coordinate along $\hat{\vec{x}}$ and $p$ is the magnetic flux per discretized ribbon (with area $N_x \Delta x \Delta y$) in units of the magnetic flux quantum $\Phi_0=h/e$. When periodic boundary conditions (PBC) are imposed, $p$ takes an integer value and represents the magnetic flux through the magnetic supercell. In general the calculations assume that the spins are independent and degenerate. For the purpose of comparing to experiment the degeneracy is relaxed by assuming an effective Land{\' e} factor with a fitted value $g=40$. In this case the eigenvalues of $H^{(\textrm{FD})}$ are shifted by $\mu_B g s_z B/\hbar$ where $\mu_B$ is the Bohr magneton and $s_z= \pm 1/2$ are the up/down components of the electron spin.

To parameterize our numerical model to our specific experimental system, we compare $R_{xx}$ of the artificial graphene heterostructure against the density of states (DOS) of the discretized system [Figs.~1e,f in the main text]. In particular, we fit the positions of the lowest-filling Landau levels (measured for $B > \SI{3}{\tesla}$) with an analytic model for an unpatterned 2D electron gas and find excellent agreement for $m^{\textrm{*}}=0.023 m_0$, which yields a Fermi level $E_F\equiv \pi n\hbar^2/m^{\textrm{*}}\sim \SI{85}{\meV}$, and $g=40$, where we are using a uniform Zeeman splitting approximation \cite{krishtopenko_theory_2011,pan_quantitative_2011}. 
The simulated energy levels are also broadened using a Lorentzian with full-width half-maximum $\SI{0.5}{\meV}$. In turn, these model parameters also yield quantitative agreement with the prominent Landau levels seen in $R_{xx}$ of our artificial graphene heterostructure.

\subsection{Efficient determination of the local Chern marker}

Numerically, the spectral localizer framework possesses two key features that lend it to efficient algorithms. First, it preserves sparsity, i.e., if $X,Y,H$ are sparse, so is $L_{(x,y,E)}$. This is in contrast to other known local Chern markers that project into the occupied subspace \cite{kitaev2006anyons,bianco_local_chern_2011}, which generally yields dense matrices. Second, the calculation of a matrix's signature does not require finding the matrix's spectrum, and can instead leverage results from applied mathematics such as Sylvester's law of inertia \cite{sylvester_xix_1852}. In particular, if $NDN^\dagger$ is the LDLT decomposition of $L_{(x,y,E)}$, then $\textrm{sig} [L_{(x,y,E)}] = \textrm{sig} \left[D \right]$. Thus, as there are fast, sparse LDLT decomposition methods \cite{duff_ma57_code_2004}, the determination of $C_{(x,y,E)}^{\textrm{L}}$ is generally orders of magnitude faster than the calculation of the spectrum of $L_{(x,y,E)}$, or finding the occupied subspace of $H$ for the same system. Overall, as the spectral localizer is built from the $K$-theory of $C^*$-algebras and is compatible with a variety of numerical speedups, it is a numerical $K$-theoretic approach to material topology.


\section{Analysis of states in between the Landau levels}

\begin{figure}[h]
    \centering
    \includegraphics{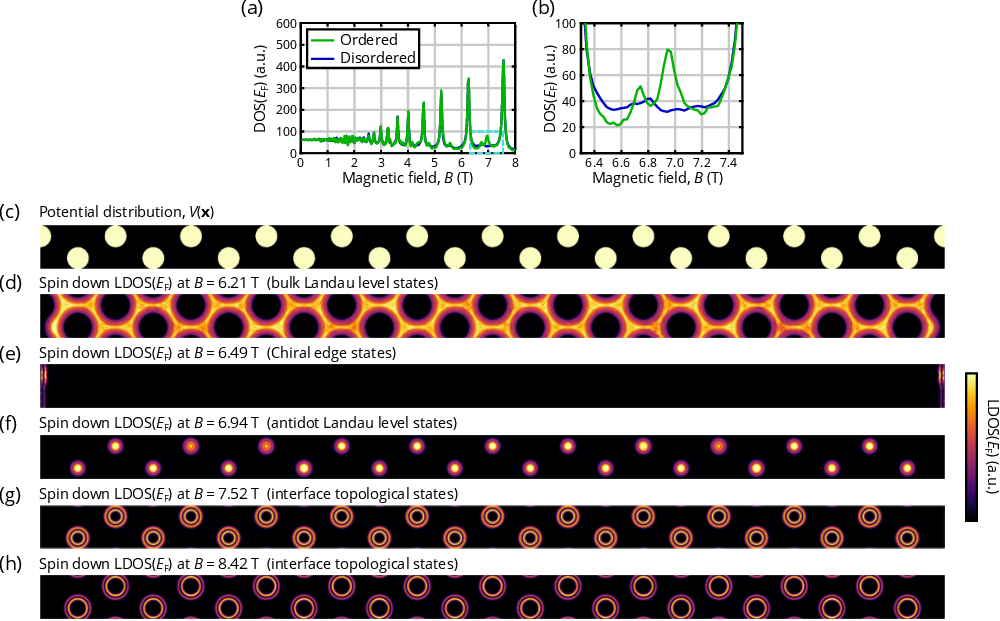}
    \caption{(a) Density of states (DOS) at the Fermi energy for a given magnetic field strength for the same simulation model as is used in Fig.~1e,f of the main text. DOS for disordered (blue, same as main text) and ordered (green) with $\delta V_\textrm{h} = \SI{0}{\meV}$ systems are shown. In other words, the data for the ordered system is the combination of Figs.~2a,b in the main text. (b) Zoom-in of the $\textrm{DOS}(E_{\textrm{F}})$ in the dashed cyan box in (a). (c-g) Real space plots of the periodic electrostatic potential (\textbf{c}), as well as the local density of states (LDOS) at the Fermi energy for the same magnetic field strengths as considered in Figs.~2c-g from the main text.}
    \label{sm_fig:dis_wfn}
\end{figure}

In the discussion of Fig.~1 of the main text, we claim that the non-zero density of states (DOS) between the Landau levels seen in simulations is due to the presence of states pinned to the antidot potential elements, whose fine features in an ordered system [main text Fig.~2] are then blurred out by disorder. In this section, we provide further evidence for this claim. In Fig.~\ref{sm_fig:dis_wfn}a,b, we compare the DOS for simulations of ordered and disordered systems for both spin sectors, and find that the inclusion of disorder blurs out the fine features seen in the ordered system in Figs.~2a,b in the main text. In Fig.~\ref{sm_fig:dis_wfn}c-h, we both show the system potential and then reproduce Figs.~2c-g from the main text on a larger scale.

\section{Verifying simulation convergence}

In this section, we demonstrate that our simulated results shown in Fig.~1e of the main text are converged with respect to the spectral localizer's scaling coefficient $\kappa$ as well as the size of the (simulated) finite flake. We also discuss the stability of the spectral localizer framework with respect to changes in $\kappa$.

\subsection{Convergence with respect to the scaling coefficient $\kappa$ \label{sec:iiiA}}

In the definition and discussion of the spectral localizer in Eq.~(2) of the main text, it is noted that the function of the scaling coefficient $\kappa$ is twofold, it ensures consistent units and it adjusts the relative weight between the position operators and Hamiltonian. This adjustment is necessary, as the two limits of $\kappa$ each yield unhelpful spectral localizers. If $\kappa = \SI{0}{\meV/\nm}$, the spectral localizer is simply the block diagonal matrix $(H-E\mathbf{1})\otimes \sigma_z$, whose signature is always zero as the two blocks have equal but opposite spectra. If $\kappa \rightarrow \infty \si{\meV/\nm}$, the spectral localizer simply contains information about the position operators, which commute, and thus $\textrm{sig}[L_{(x,y,E)}] = 0$ as well. (Further discussion of these two limits of $\kappa$ is provided in Sec.~4A of Ref.~\cite{cerjan_classifying_2024}.) As such, the spectral localizer only provides a correct determination for material topology for a range of intermediate values of $\kappa$ between these two limits, when the spectral weight between the Hamiltonian and position operators in $L_{(x,y,E)}$ is relatively balanced.

\begin{figure}[h]
    \centering
    \includegraphics{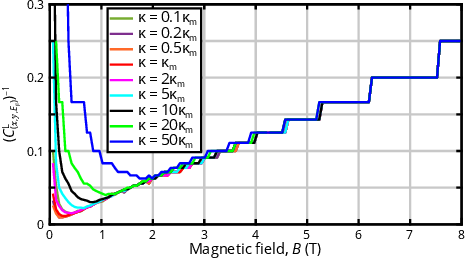}
    \caption{Inverse of the local Chern marker at the Fermi energy versus magnetic field calculated for the same model parameters as Fig.~1e in the main text, using different values of $\kappa$. Here, $\kappa_\textrm{m} = \SI{1E-3}{\meV/\nm}$ is the value used in the main text.}
    \label{sm_fig:kappa}
\end{figure}

For tight-binding models (or other models featuring bounded Hamiltonians) that describe single materials characterized by a single length scale, a range of validity for $\kappa$ can be proven \cite[Theorem 2]{LoringSchuBa_even}. However, one of these bounds depends on the $\ell^2$ norm of the Hamiltonian; thus, for unbounded systems this bound on $\kappa$ is not definable, as the $\ell^2$ norm is not defined. It is not currently known how to adapt these proven bounds to the case of unbounded Hamiltonians. Nevertheless, in every case that we are aware of, the spectral localizer for unbounded systems still exhibits a range of $\kappa$ over which the topological invariant is stable \cite{cerjan_operator_Maxwell_2022,dixon_classifying_2023,cerjan_local_crystal_marker_2024}, generally near $\kappa \sim E_{\textrm{gap}}/L_{\textrm{min}}$ as discussed in the main text. Indeed, Fig.~\ref{sm_fig:kappa} shows that the results in Fig.~1e of the main text can be quantitatively reproduced at larger magnetic fields for a range of $\kappa$ spanning more than two orders of magnitude.

In our simulations, we are usually choosing $\kappa \sim E_{\textrm{gap}}/L_{\textrm{min}}$, where $L_{\textrm{min}}$ is the smallest dimension of the finite system and $E_{\textrm{gap}} = \hbar \omega_\textrm{c}$ is the width of the relevant bulk spectral gap based on the cyclotron frequency at $B = \SI{7}{\tesla}$. As such, it is expected that the local Chern number will show better convergence with respect to $\kappa$ for larger magnetic fields, as discussed in the main text. We expect that the spectral localizer framework can accurately resolve the local Chern marker at smaller magnetic fields using a different $\kappa$ and a larger simulation domain to increase $L_{\textrm{min}}$. 

However, for multi-scale systems that can exhibit different properties at different length scales, changing $\kappa$ enables the spectral localizer framework to be sensitive to the different topological phenomena at these different length scales. For example, while Fig.~1 in the main text analyzes the number of chiral edge states that appear at the boundaries of the sample, Fig.~2 studies the origins of the states pinned to each antidot that are dependent on the potential strength and diameter of the antidot and not the full sample. Thus, Fig.~1 uses $\kappa \sim E_{\textrm{gap}}/L_{\textrm{min}}$, while Fig.~2 uses the larger $\kappa \sim 2 E_{\textrm{gap}}/a$, where $a/2$ is the antidot diameter. Mathematically, the increased $\kappa$ is placing greater emphasis on the position operators in $L_{(x,y,E)}$, so the signature is more responsive to changes in $(x,y)$ and less responsive to changes in $E$, effectively limiting the function of the spectral localizer to larger magnetic fields where the local spectral gap is larger.

On its surface, the ability to change $\kappa$ and obtain a different topological marker appears to present some confusion --- for a given choice of $(x,y,E)$, what is the system's actual topology? For example, if $(x,y)$ are chosen in the center of an antidot of artificial graphene and $E = E_{\textrm{F}}$, the system appears to provide $3$ different values for the local Chern marker. However, these different values are not actually in conflict, they simply reveal different physics on different length scales. For $\kappa \sim E_{\textrm{gap}}/L_{\textrm{min}}$, the local Chern marker is revealing sample-scale topology, and will correspond to the number of chiral edge states around the boundary of the full system. For $\kappa \sim 2 E_{\textrm{gap}}/a$, the local Chern marker is showing antidot-scale topology, and predicts the number of antidot-bulk interface localized states. For $\kappa \ll E_{\textrm{gap}}/L_{\textrm{min}}$, the local Chern marker is $0$, because there is no relevant physics at much larger length scales nor any material boundary or interface to support a boundary-localized state. Likewise, for $\kappa \gg 2 E_{\textrm{gap}}/a$, the local Chern marker is again $0$, again due to the lack of relevant phenomena at much shorter length scales.

In principle, the $\kappa$ coefficient in front of $(X-x\mathbf{1})$ and $(Y-y\mathbf{1})$ could be different. However, as the scale of both position operators is set by the lattice constant $a$, choosing different $\kappa$ adds unnecessary complexity for little benefit. If these two position operators were set by different system length scales, using different $\kappa$ would likely be required.

\subsection{Stability of the spectral localizer framework's predictions with respect to $\kappa$}

Broadly, the dependence of the spectral localizer framework on a scaling parameter such as $\kappa$ is both expected, and can be viewed as a feature rather than a bug. First, to the best of our knowledge, all other frameworks for local topological markers have an equivalent choice of hyper-parameter. For example, both the Kitaev \cite[Appendix C]{kitaev2006anyons} and Bianco-Resta \cite{bianco_local_chern_2011} markers require a choice of integration region(s) that must be sufficiently large so as to guarantee an approximately integer-valued local marker, but not so large as to contain any portion of a material's boundary (if a material's full boundary is contained, both of these local markers are provably $0$). In the spectral localizer framework, $\kappa$ is fulfilling a similar purpose as the choice of integration region in these other frameworks, with larger values of $\kappa$ approximately corresponding to smaller integration regions that yield greater spatial specificity in the topological classification of a material at the cost of reduced spectral accuracy. Note, the spectral localizer's markers are always integer-valued regardless of the choice of $\kappa$.

Moreover, as discussed in Sec.~\ref{sec:iiiA}, the ability to change $\kappa$ allows the spectral localizer framework to both resolve sample-scale topology as well as internal regions within a sample that may have different local topology, but that do not contribute to a standard edge-conductance measurement (e.g., the different local topology of the antidots discussed in the main text in Fig.~2). But, how does one know when to trust that a calculated value of the local marker is revealing material phenomena rather than numerical noise? In this section, we detail how the spectral localizer framework comes inherently equipped with a measure of topological protection that can be heuristically viewed as a local spectral gap. This measure of robustness also includes protection against changes in $\kappa$. Thus, when this measure of protection is large, relative to the energy scales of the system, the spectral localizer's predictions correspond to real physical phenomena.

\begin{figure}[th]
    \centering
    \includegraphics{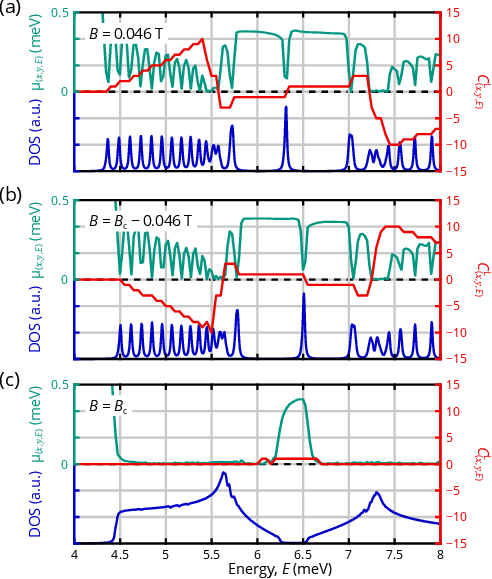}
    \caption{(a-c) Local gap $\mu_{(x,y,E)}$ (green), density of states (blue) and $C_{(x,y,E)}^{\textrm{L}}$ in the unpatterned bulk (red) for a single spin sector of artificial graphene with $V_{\textrm{h}} \rightarrow \infty \si{\meV}$ using the same simulation parameters as Fig.~3 in the main text for magnetic field strengths near $B \sim \SI{0}{\tesla}$ and the critical field $B \sim B_\textrm{c}$ where $\Phi = \Phi_0$.}
    \label{sm_fig:mu}
\end{figure}

First, mathematically observe that changes in $\kappa$ always change the spectrum of $L_{(x,y,E)}$ continuously and thus the local marker $C_{(x,y,E)}^{\textrm{L}}$ is necessarily stable against changes in $\kappa$ so long as $L_{(x,y,E)}$ remains invertible. In particular, so long as $H$ is bounded, $L_{(x,y,E)}$ is Hermitian and bounded, and as such its eigenvalues must move continuously with respect to perturbations to $L_{(x,y,E)}$ at a rate limited by Weyl's inequality \cite{weyl_asymptotische_1912,Bhatia1997}. For the specific case of perturbations in $\kappa$, if $\kappa \rightarrow \kappa + \delta \kappa$, then $L_{(x,y,E)} \rightarrow L_{(x,y,E)} + \delta L_{(x,y,E)}$ with
\begin{equation}
    \delta L_{(x,y,E)} = \delta \kappa (X-x{\bf 1})\otimes \sigma_x + \delta \kappa (Y-y{\bf 1})\otimes \sigma_y,
\end{equation}
and thus if $\lambda_j^{(\kappa)} \in \textrm{spec}[L_{(x,y,E)}]$ is the $j$th smallest eigenvalue of $L_{(x,y,E)}$, and $\lambda_j^{(\kappa+\delta \kappa)} \in \textrm{spec}[L_{(x,y,E)} + \delta L_{(x,y,E)}]$ is similarly defined, then by Weyl's inequality
\begin{equation}
    \left \vert \lambda_j^{(\kappa+\delta \kappa)} - \lambda_j^{(\kappa)} \right \vert \le \left \Vert \delta L_{(x,y,E)} \right \Vert. \label{sm_eq:weyl}
\end{equation}
Here, $\textrm{spec}[M]$ denotes the spectrum of $M$. Now, as the local Chern marker $C_{(x,y,E)}^{\textrm{L}}$ defined in Eq.~3 of the main text is given by the difference in the number of positive eigenvalues minus the number of negative ones, the only way for $C_{(x,y,E)}^{\textrm{L}}$ to change its value is if one of the spectral localizer's eigenvalues first becomes $0$, i.e., $L_{(x,y,E)}$ becomes non-invertible. Altogether, this means that the spectral localizer framework is inherently equipped with a quantitative measure of topological protection
\begin{equation}
    \mu_{(x,y,E)}(X,Y,H) = \min\left( \left\vert \textrm{spec}\left[L_{(x,y,E)}(X,Y,H) \right] \right\vert \right), \label{sm_eq:mu}
\end{equation}
in other words, the smallest distance an eigenvalue of the spectral localizer must move before the system's local topological marker can change. Any perturbation to the spectral localizer with $\Vert \delta L_{(x,y,E)} \Vert < \mu_{(x,y,E)}$ cannot change the local Chern marker. This eigenvalue movement could be due to a chance in choice of $(x,y,E)$, a perturbation to the physical system $H \rightarrow H + \delta H$, or a change in $\kappa$ as discussed. An example of this measure of topological protection is shown in Fig.~\ref{sm_fig:mu}, where $\mu_{(x,y,E)}$ is plotted for changes in energy corresponding to the unconventional Landau levels discussed in Fig.~4 of the main text. As can be seen, the system's topological marker can only change across regions where $\mu_{(x,y,E)} \rightarrow 0$. (Note, the lack of exact zeros in Fig.~\ref{sm_fig:mu} is simply due to the numerical sampling not perfectly aligning with the energies where these closings actually occur.)

Altogether, the predictions of the spectral localizer framework for a given $\kappa$ can be trusted when the local gap $\mu_{(x,y,E)}$ is similar to or larger than the relevant energy scale of the system. (Note that by definition $\mu_{(x,y,E)}$ has units of energy.)
In practice, one generally wants to coarsely sweep $\kappa$ over a few orders of magnitude in the vicinity of $\kappa \sim E_{\textrm{gap}}/L_{\textrm{min}}$ for choices of $(x,y,E)$ where one suspects a system might be topological and choices of $L_{\textrm{min}}$ corresponding to a system's relevant length scales. While this sweep may miss small $\kappa$ intervals where the system has a non-trivial local marker, by Eqs.~\eqref{sm_eq:weyl} and \eqref{sm_eq:mu} the robustness of any associated topological phenomena will necessarily be weak. This weakness could manifest as either an associated edge state being sensitive to system perturbations, or a difficulty in even identifying an associated edge state(s) in the system's LDOS as the state is delocalized; small values of $\mu_{(x,y,E)}$ correspond to locations where the system has an approximate state \cite{cerjan_quadratic_2022} so large regions where $\mu_{(x,y,E)} \approx 0$ typically indicate regions with one or many delocalized states (e.g., as can be seen via comparison of $\mu_{(x,y,E)}$ with the DOS in Fig.~\ref{sm_fig:mu}).

Before concluding this section, we note that recent mathematical results show that the entire operation of the spectral localizer framework for classifying 1D systems can be understood through the response of the spectrum of $L_{(x,y,E)}$ to changes in $\kappa$ \cite{shapiro_loringschulz-baldes_2025}, demonstrating that $\kappa$ may be even more integral to the spectral localizer framework than previously appreciated.

\subsection{Convergence with respect to simulation domain size}

\begin{figure}[h]
    \centering
    \includegraphics{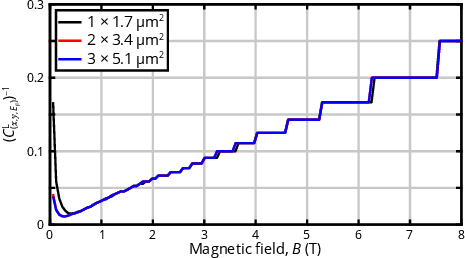}
    \caption{Inverse of the local Chern marker at the Fermi energy versus magnetic field calculated for the same model parameters as Fig.~1e in the main text, using different simulation domains $\sim1\times1.7~\SI{}{\micro\meter^2}$, $\sim2\times3.4~\SI{}{\micro\meter^2}$ (same as main text), and $\sim3\times5.1~\SI{}{\micro\meter^2}$.}
    \label{sm_fig:size}
\end{figure}

As the spectral localizer must be used in conjunction with a finite system, here we confirm that the results shown in Fig.~1e for the local Chern marker are converged with respect to the choice of simulated system size. In Fig.~\ref{sm_fig:size} we show calculations using three different system sizes that are nearly identical, especially in the high-field limit, despite featuring very different system sizes.

\section{Band structures for the emergence of Hofstadter's butterfly}

\begin{figure}[t]
    \centering
    \includegraphics{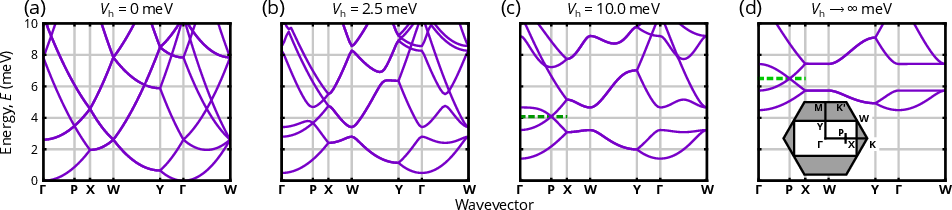}
    \caption{
    (a-d) Artificial graphene's bulk band structures at $B = \SI{0}{\tesla}$ for $V_{\textrm{h}} = \SI{0}{\meV}$ (a) to $\SI{2.5}{\meV}$ (b), $\SI{10}{\meV}$ (c), and finally $\infty \si{\meV}$ (d). The lowest energy Dirac point is marked in the systems with spectrally isolated Dirac points (green dashed lines), and this energy is indicated in the corresponding Chern marker plots (green arrows) in Fig.~3 of the main text.
    }
\label{fig:L80Bands}
\end{figure}

In Fig.~3 of the main text, we show the emergence of Hofstadter's butterfly as the strength of the periodic potential is increased from zero. In Fig.~\ref{fig:L80Bands}, we show the corresponding bulk band structures at $B = \SI{0}{\tesla}$ to show the energy of the Dirac point as well as the energy of the bottom of the bands.

\section{Hofstadter's butterfly at larger scales}

\begin{figure}[tb]
    \centering
    \includegraphics{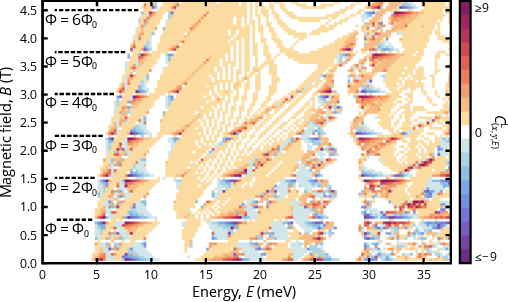}
    \caption{Predicted local Chern marker $C_{(x,y,E)}^{\textrm{L}}$ in the unpatterned bulk for a single spin sector as a function of magnetic field strength and energy for the same ordered artificial graphene system considered in Fig.~3 of the main text with $V_{\textrm{h}} \rightarrow \infty \si{\meV}$.}
\label{fig:big}
\end{figure}

Compared to natural graphene, artificial graphene offers broad versatility as its properties can be modified by adjusting the parameters of the nanoscale antidot pattern, such as the lattice constant and antidot potential strength. However, in contrast to natural graphene, probing the energy range near the Dirac point can be challenging in artificial graphene, as achieving $E_\textrm{F} \sim E_{\textrm{D}}$ requires reaching electron densities on the order of $n \sim \SI{e10}{\per \cm\squared}$ for typical lattice periodicities of \SI{100}{\nano\meter}. Nevertheless, the spectral localizer reveals that it is possible to observe Hofstadter phenomena, such as the rapid sign change of the Chern marker across lines where $\Phi/\Phi_0 \in \mathbb{Z}$ and potentially even full butterflies, for $E_\textrm{F} \gg E_{\textrm{D}}$ in artificial graphene [Fig.~\ref{fig:big}]. Here, the Chern marker fringes seen for stronger $B$ are likely a finite size effect, and can be qualitatively reproduced in tight-binding simulations of a honeycomb lattice [see Sec.~\ref{sec:s5}]. Thus, so long as an experimental system can realize a sufficiently strong antidot potential $V_\textrm{h} > E_\textrm{F}$, the spectral localizer predicts that Hofstadter phenomena should manifest for larger electron densities that are easier to obtain in experiments; for example, rapid changes in the Chern marker appear in Fig.~\ref{fig:big} for $E_\textrm{F} = \SI{35}{\meV}$, which corresponds to $n \sim \SI{4.4e11}{\per \cm\squared}$.

\subsection{Local Chern marker fringes in tight binding models \label{sec:s5}}

\begin{figure}[h]
    \centering
    \includegraphics{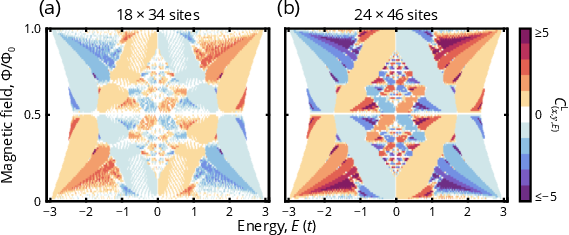}
    \caption{Local Chern marker $C_{(x,y,E)}^{\textrm{L}}$ in the unpatterned bulk as a function of energy and magnetic field strength in normalized units for a tight-binding honeycomb lattice with $18\times34$ (a) or $24\times46$ (b) sites. Simulations use $\kappa = 0.007 (t/a)$ where $t$ is the nearest neighbor coupling strength and $a$ is the lattice constant.}
    \label{sm_fig:fringes}
\end{figure}

In Fig.~\ref{fig:big}, simulations of the local Chern marker for artificial graphene, fringes are seen for larger energies and magnetic field strengths. In this section, we demonstrate that these fringes are consistent with finite size effects seen in calculations of the local Chern marker in a tight-binding lattice. Specifically, we consider a tight-binding honeycomb lattice with only nearest neighbor couplings \cite{agazzi_colored_2014}, and calculate its local Chern marker using the spectral localizer for two different choices of finite flakes. As can be seen in Fig.~\ref{sm_fig:fringes}, for small simulation domains, fringes appear in some of the butterfly's minigaps that should show a single value for the local Chern marker. As the simulation domain is increased in size, these fringes disappear. Additionally, the larger simulation domain enables the resolution of local Chern markers with larger magnitudes. This latter behavior is consistent with the difference seen between Fig.~1c and Fig.~1e of the main text between the local Chern marker and measured Hall conductivity at low magnetic field strengths; resolving larger Chern marker magnitudes requires ensuring the system is large enough so that the corresponding edge-localized states do not couple to states from the opposite edge, rendering the states trivial. Similarly, we suspect that the fringes in regions with small-magnitude local Chern markers originate from the angular quantization of the chiral edge states due to the simulated system's finite size---the smaller the system area, and thus the smaller the system boundary, the larger the spectral spacing between the edge states, potentially providing an opportunity for the spectral localizer to misclassify the system's topology within these spectral gaps. Altogether, we suspect that the fringes seen in Fig.~\ref{fig:big} of the main text could also be returned to a single Chern marker by increasing the simulation domain of the artificial graphene used.

\begin{figure}[t]
    \centering
    \includegraphics{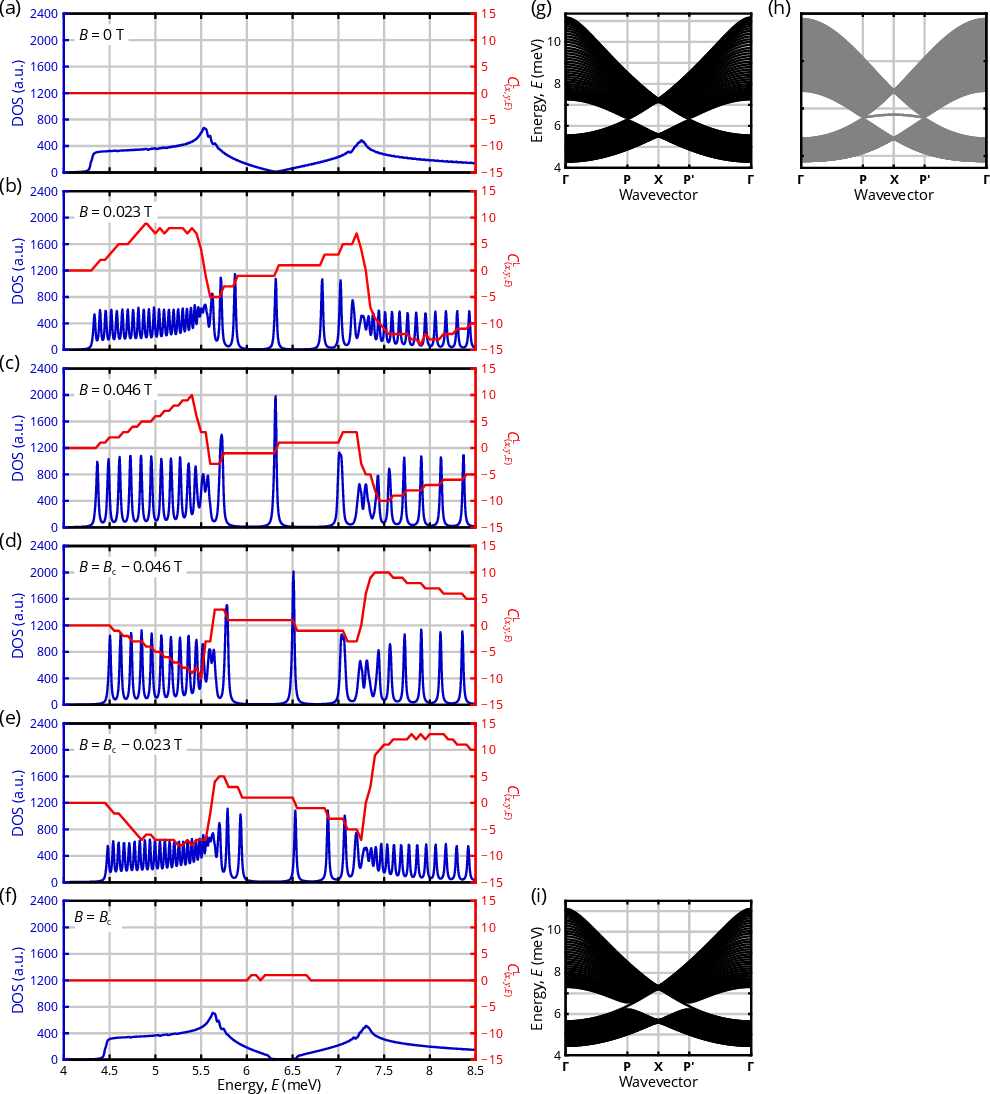}
    \caption{(a-f) Density of states (blue) and $C_{(x,y,E)}^{\textrm{L}}$ in the unpatterned bulk (red) for a single spin sector of artificial graphene with $V_{\textrm{h}} \rightarrow \infty \si{\meV}$ using the same simulation parameters as Fig.~3 in the main text for magnetic field strengths near $B \sim \SI{0}{\tesla}$ and the critical field $B \sim B_\textrm{c}$ where $\Phi = \Phi_0$. Some panels show the same data as Fig.~4 in the main text, except using a uniform DOS axis to allow for direct comparison. (g-i) Ribbon band structures along the remaining wavevector with open boundary conditions in the other direction for the continuum model at $B = \SI{0}{\tesla}$ (g), a honeycomb tight-binding lattice with $t_{\textrm{NNN}}/t_{\textrm{NN}} = 0.13$ (h), and the continuum model at $B = B_\textrm{c}$ with a convenient choice of gauge for the vector potential (i).}
    \label{sm_fig:AGcuts}
\end{figure}

\section{Extended data for the unconventional quantum Hall effect in artificial graphene}

In Fig.~4 of the main text, we discuss the appearance of the unconventional quantum Hall effect in artificial graphene near both $B = \SI{0}{\tesla}$ and $B = B_\textrm{c}$. In Fig.~\ref{sm_fig:AGcuts}a-f, we show additional data to help verify both the square root dependence of the Landau level gaps as well as the change in Chern number by $2$ per spin sector. Moreover, we also compare the ribbon band structures (i.e., the band structures of systems that are periodic in one direction and open in the other) of the continuum models at $B = \SI{0}{\tesla}$ and $B = B_\textrm{c}$, as well as a tight-binding model with with nearest neighbor (NN) and next-nearest neighbor (NNN) couplings $t_{\textrm{NNN}}/t_{\textrm{NN}} = 0.13$. First, by comparing Fig.~\ref{sm_fig:AGcuts}g and Fig.~\ref{sm_fig:AGcuts}h, we conclude that artificial graphene has next-nearest neighbor couplings between its pseudo-atoms that are $\sim 13\%$ of the strength of the couplings between the nearest neighbor pseudo-atoms, but also that there are additional longer-range couplings beyond NNN couplings in artificial graphene, as the agreement is not perfect. We note that the difference between the edge states along the direction with open boundaries is due to artificial graphene having a half-potential at its edge [see Fig.~\ref{sm_fig:dis_wfn}c] that is known to change the dispersion of the edge states \cite{yao_edge_2009}. Similarly, comparing Fig.~\ref{sm_fig:AGcuts}g and Fig.~\ref{sm_fig:AGcuts}i shows the opening of a bulk band gap crossed by a pair of chiral edge states, confirming that artificial graphene becomes a topological insulator at $B = B_\textrm{c}$.

\bibliography{predicting_butterfly_refs}